\documentclass{article}

\usepackage{arxiv}

\usepackage[utf8]{inputenc} 
\usepackage[T1]{fontenc}    
\usepackage[
    colorlinks=true,
    linkcolor=black,
    citecolor=black,
    filecolor=black,
    urlcolor=black,
]{hyperref}                 
\usepackage{url}            
\usepackage{booktabs}       
\usepackage{amsfonts}       
\usepackage{nicefrac}       
\usepackage{microtype}      
\usepackage{lipsum}		
\usepackage{graphicx}
\usepackage{svg}
\usepackage[numbers,sort&compress]{natbib}
\usepackage{doi}

\renewcommand{\b}[1]{\textbf{#1}}

\title{Predicting small molecules solubility on endpoint devices using deep ensemble neural networks}
\date{\today}
\author{
    \href{https://orcid.org/0000-0001-5336-2847}{\includegraphics[scale=0.06]{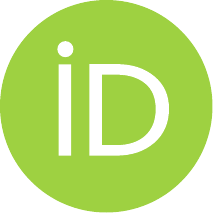}\hspace{1mm}Mayk Caldas Ramos}\\
	Department of Chemical Engineering \\
	University of Rochester\\
	\texttt{mcaldasr@ur.rochester.edu} \\
	\And
	\href{https://orcid.org/0000-0002-6647-3965}{\includegraphics[scale=0.06]{orcid.pdf}\hspace{1mm}Andrew D. White} \\
	Department of Chemical Engineering \\
	University of Rochester\\
	\texttt{andrew.white@rochester.edu} \\
}

\renewcommand{\shorttitle}{Predicting small molecules solubility using deep ensemble neural networks}

\hypersetup{
pdftitle={Predicting small molecules solubilities on endpoint devices using deep ensemble neural networks},
pdfsubject={},
pdfauthor={Mayk C.~Ramos, Andrew D.~White},
pdfkeywords={Solubility, Small Molecule, Deep Ensemble, Recurrent Neural Network},
}

\begin{document}
\maketitle

\begin{abstract}
Aqueous solubility is a valuable yet challenging property to predict.
Computing solubility using first-principles methods requires accounting for the competing effects of entropy and enthalpy, resulting in long computations for relatively poor accuracy. 
Data-driven approaches, such as deep learning, offer improved accuracy and computational efficiency but typically lack uncertainty quantification.
Additionally, ease of use remains a concern for any computational technique, resulting in the sustained popularity of group-based contribution methods.
In this work, we addressed these problems with a deep learning model with predictive uncertainty that runs on a static website (without a server). 
This approach moves computing needs onto the website visitor without requiring installation, removing the need to pay for and maintain servers. 
Our model achieves satisfactory results in solubility prediction.
Furthermore, we demonstrate how to create molecular property prediction models that balance uncertainty and ease of use. 
The code is available at \url{https://github.com/ur-whitelab/mol.dev}, and the model is usable at \url{https://mol.dev}.
\end{abstract}

\section{Introduction}

Aqueous solubility measures the maximum quantity of matter that can be dissolved in a given volume of water.
It depends on several conditions, such as temperature, pressure, pH, and the physicochemical properties of the compound being solvated.\cite{Sorkun2019-jv}
The solubility of molecules is essential in many chemistry-related fields, including drug development\cite{Dajas2012-jg, Di2012-wd, Docherty2015-pc, Barrett2022-hr}, protein design\cite{Sormanni2015-js}, chemical\cite{Herrero-Martinez2005-kl, Diorazio2016-dn} and separation\cite{Sheikholeslamzadeh2012-wq} processes.
In drug development, for instance, compounds with biological activity may not have enough bioavailability due to inadequate aqueous solubility.

Solubility prediction is essential, driving the development of various methods, from physics-based approaches — including first principles\cite{Yalkowsky1980-qx, Ran2001-dn}, semi-empirical equations\cite{Fredenslund1975-xb, Abrams1975-ar, Maurer1978-vg}, molecular dynamics (MD)\cite{Luder2007-mw, Luder2007-pp, Boothroyd2018-lo, Boothroyd2019-ys}, and quantum computations\cite{Tomasi2005-bf} — to empirical methods like quantitative structure-property relationship (QSPR)\cite{Yu2006-lg, Ghasemi2007-cc, Duchowicz2009-kq, Louis2009-xi} and multiple linear regression (MLR)\cite{Huuskonen2000-pb, Delaney2004-de}. Despite their sophistication, physics-based models often present complexity that limits accessibility to advanced users\cite{Skyner2015-pz} and do not guarantee higher accuracy than empirical methods\cite{McDonagh2014-bj}. Data-driven models emerge as efficient alternatives, capable of outperforming physics-based models\cite{Skyner2015-pz}. However, achieving accurate and reliable solubility predictions remains a significant challenge\cite{Skyner2015-pz, Sorkun2021-kc}.

To address the persistent issues of systematic bias and non-reproducibility in aqueous solubility datasets, Llinàs et al.\cite{Llinas2008-rc, Llinas2019-eu} introduced two solubility challenges featuring consistent data. The first challenge red participants based on the root mean square error (RMSE) and the accuracy within a $\pm0.5$ logS error range\cite{Hopfinger2009-rt}. The second challenge revealed that despite the freedom in method selection, all entries relied on QSPR or machine learning (ML) techniques\cite{Llinas2020-ea}, yet did not achieve a notable improvement over the first challenge\cite{Hopfinger2009-rt}. These challenges highlighted the importance of data quality over model selection for accurate solubility predictions.\cite{Llinas2020-ea}
\citet{Sorkun2021-kc} further emphasized this by demonstrating how data quality assessments on subsets of the AqSolDB\cite{Sorkun2019-jv} significantly impacted model performance\cite{Sorkun2021-kc}.

\citet{McDonagh2014-bj} demonstrated that cheminformatic methods surpass first principle theoretical calculations in calculating solubilization free energies, highlighting the superior accuracy of cheminformatics and the efficacy of Random Forest models, evidenced by an RMSE of 0.93 using Llinàs' first dataset\cite{Llinas2008-rc}.
Data-driven approaches, particularly feature-based models, have contributed to accurate aqueous solubility prediction.
\citet{Delaney2004-de} used MLR to develop a model called Estimated SOLubility (ESOL) adjusted on a 2874 small organic molecules dataset with an average absolute error (AAE) of 0.83.
Comparable performance has been achieved using various methods including MLR\cite{Huuskonen2000-pb}, Gaussian processes\cite{Schwaighofer2007-bf}, undirected graph recurrent neural networks (UG-RNN)\cite{Lusci2013-rp}, deep neural networks (DNN)\cite{Ye2021-xv}, and random forests (RF)\cite{Tayyebi2023-ad, Kurotani2021-gb}.

Recently, transformers\cite{Vaswani2017-dj} models have been applied to compute solubility of small molecules.\cite{Wang2019-lm, Fabian2020-vr, Francoeur2021-zr, Born2022-yt, Ross2022-na}.
\citet{Francoeur2021-zr} developed the SolTranNet, a transformers model trained on AqSolDB\cite{Sorkun2019-jv} solubility data.
Notably, this architecture results in an RMSE of only $0.278$ when trained and evaluated on the original ESOL\cite{Delaney2004-de} dataset using random split.
Nevertheless, it shows an RMSE of $2.99$ when trained using the AqSolDB\cite{Sorkun2019-jv} and evaluated using ESOL.
It suggests that the molecules present in ESOL may have low variability, meaning that samples in the test set are similar to samples in the training set.
Hence, models trained on the ESOL training set performed excellently when evaluated on the ESOL test set. 

Solubility models should ideally combine accuracy with ease of access.
Thus, a common idea is to use web servers to provide easier public access.
However, web servers demand continuous financial and time investments for maintenance, leading to the eventual disappearance of some, despite having institutional or government backing\cite{zdrazil2017rise}. For instance, eight out of 89 web server tools featured in the 2020 \textit{Nucleic Acids Research} special web server issue were offline by the end of 2022\cite{seelow2020editorial}. 
Moreover, computational demands can be significant, with tools like RoseTTAFold\cite{Baek2021-at} and ATB\cite{Stroet2018-ag} requiring hours to days for job completion, thus creating potential delays due to long queues and wait times\cite{Smith2021-re}.

An alternative approach is to perform the computation directly on the user's device, removing the need for the server's maintenance and cost. 
This method allows hosting the website as a static file on platforms such as GitHub, with potential archiving on the Internet Archive\footnote{https://archive.org/}.
We explored this approach in \citet{Ansari2023} for bioinformatics.
Our web application implements a deep ensemble\cite{Lakshminarayanan2016-ck} recurrent neural network (RNN) capable of extracting data directly from molecular string representations, such as SMILES\cite{Weininger1988-lh} or SELFIES\cite{Krenn2022-re}, which can be easily quickly accessed.\cite{Kim2018-kb, Schilter2024-oe}

The primary difficulty lies in the application's dependence on the device's capabilities, which is crucial for smartphones with limited resources. Balancing performance in low-resource settings, the use of transformer models\cite{Vaswani2017-dj} becomes impractical due to their large size, incompatible with smartphone memory and prolonged inference times. Additionally, our model implements a deep ensemble to calibrate uncertainties, making the application of transformers even more unfeasible.
In contrast, using descriptors is an easy way to convey physical information to the model and, consequently, enables smaller models. However, descriptor computation is time-intensive. In our tests, using PaDEL to compute descriptors for all molecules in AqSolDB took roughly $\sim$~20 hours. Furthermore, feature-based model development requires specialized knowledge for feature selection\cite{Beltran2018-we}, and is limited by the regions of the chemical space these descriptors cover\cite{maggiora2006outliers}.
Even application usage may need specialized data, as \citet{Kurotani2021-gb} illustrate. 
RNNs present an alternative for property extraction directly from string representations while allowing for adaptable computational resource management.

In this work, we developed a front-end application using a JavaScript (JS) implementation of TensorFlow framework\cite{Smilkov2019-ye}.
Our application can be used to predict the solubility of small molecules with uncertainty.
To calibrate the confidence of the prediction, our model implements a deep ensemble approach\cite{Lakshminarayanan2016-ck} which allows reporting model uncertainty when reporting the prediction.
Our solution implements a deep ensemble of RNN models specially designed to achieve satisfactory performance while being able to run in an environment without strong computational resources.
This application runs locally on the user's device and can be accessed at \url{https://mol.dev/}.
Mol.dev does not save data input for predictions in any way.

\section{Methods}
\label{sec:Methods}

\subsection{Dataset}

The data used for training the models were obtained from AqSolDB\cite{Sorkun2019-jv}. 
This database combined and curated data from 9 different aqueous solubility datasets.
The main concern in using a large, curated database is to avoid problems with the generalizability of the model\cite{Wang2009-fc} and with the fidelity of the data\cite{Wang2011-lk}. 
AqSolDB consists of aqueous solubility (LogS) values for 9982 unique molecules extended with 17 topological and physicochemical 2D descriptors calculated by RDKit\cite{Landrum_undated-nt}.

We augmented AqSolDB to 96,625 molecules using SMILES randomization\cite{Arus-Pous2019-nl, Schwaller2020-lw}.
Each entry of AqSolDB was used to generate at most ten new unique randomized SMILES strings.
Training the model on multiple representations of the same molecule improves its ability to learn the chemical space constraints of the training set, as demonstrated in previous studies \cite{Arus-Pous2019-nl, Schwaller2020-lw}.
Duplicates were removed.

After shuffling, the augmented dataset was split into 80\%/20\% for the training and test datasets, respectively.
The curated datasets for the solubility challenges\cite{Llinas2008-rc, Llinas2020-ea} were used as withheld validation data to evaluate the model's ability to predict solubility for unseen compounds.
To refer to the validation datasets, we labeled the first solubility challenge dataset as "solubility challenge 1" and the two sets from the second solubility challenge as "solubility challenge 2\_1" and "solubility challenge 2\_2", respectively. 
Molecules in these three datasets were not found in train and test datasets.

\subsection{Model architecture}


Our model uses a deep ensemble approach as described by \citet{Lakshminarayanan2016-ck}.
This technique was selected due to its ability to estimate prediction uncertainty, thus enhancing the predictive capability of our model. 
The uncertainty of a model can be divided into two sources: aleatoric uncertainty (AU) and epistemic uncertainty (EU).\cite{Shaker2020-gz, Ghoshal2021-pi}
These uncertainties quantify the intrinsic uncertainty inherent in data observations and the disagreement among model estimations, respectively.\cite{Scalia2020-kq}

Given a model that outputs two values -- $\hat{\mu}_m$ and $\hat{\sigma}_m$ -- that characterize a normal distribution $\mathcal{N}(\hat{\mu}_m,\hat{\sigma}_m)$, a deep ensemble creates an ensemble of $m$ models that can estimate prediction uncertainty.
For a given data point $\vec{x}$, the estimates for the ensemble predictions are computed as follows:

\begin{equation}
    \hat{\mu}(\vec{x}) = \frac{1}{N}\sum_m \hat{\mu}_m(\vec{x})
\end{equation}

\begin{equation}
    \hat{\sigma}^2_{ale}(\vec{x}) = \frac{1}{N}\sum_m \hat{\sigma}^2_m(\vec{x}) \ , \ \ \hat{\sigma}^2_{epi}(\vec{x}) = \frac{1}{N}\sum_m \left(\hat{\mu}(\vec{x}) - \hat{\mu}_m(\vec{x}) \right)^2
\end{equation}
where $\hat{\sigma}^2_{ale}$ is AU, $\hat{\sigma}^2_{epi}$ is EU, N is the ensemble size, and m indexes the models in the ensemble.

\begin{figure}[ht]
    \centering
    \includegraphics[width=\textwidth]{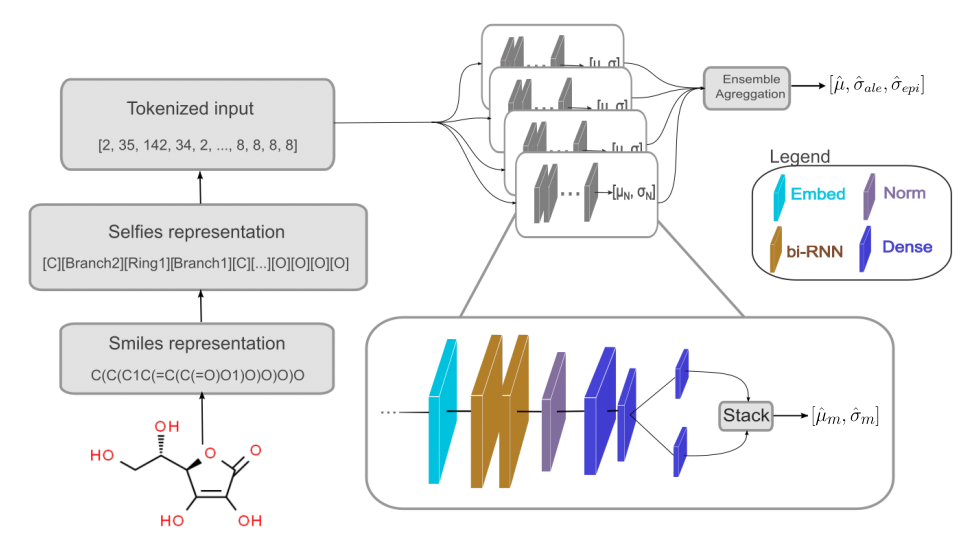}
    \caption{Scheme of the deep learning DNN. 
    The molecule is input using the SMILES or SELFIES representation. 
    This representation is converted to a tokenized input based on a vocabulary obtained using the training dataset.
    A set of models represents the deep ensemble model.
    Each model consists of an embed layer, two bidirectional RNN (bi-RNN) layers, a normalization layer, and three fully connected layers being down-sized in three steps.
    Dropout layers are present after the embed and after each fully connected layer during training, but they were not represented in this scheme.
    Predictions of the models in the ensemble are then aggregated.}
    \label{fig:model}
\end{figure}

We used a deep neural network (DNN) implemented using Keras\cite{Chollet2018-vq} and TensorFlow\cite{Abadi2015-xq} to build the deep ensemble.
Our DNN model uses Self-referencing embedded strings (SELFIES)\cite{Krenn2022-re} tokens as input.
A pre-defined vocabulary was created by analyzing all training data. Each unique SELFIES group was assigned to a corresponding integer, yielding 273 distinct tokens.
Simplified molecular-input line-entry system (SMILES)\cite{Weininger1988-lh} or SELFIES\cite{Krenn2022-re} molecule representations are converted to tokens based on the pre-defined vocabulary.
Figure \ref{fig:model} illustrates the model architecture.
The network can be divided into three sections:
($i$) Embedding,
($ii$) bi-RNN, and
($iii$) fully connected NN.

The embedding layer converts a list of discrete tokens into a fixed-length vector space.
Working on a continuous vector space has two main advantages: it uses a more compact representation, and semantically similar symbols can be described closely in vector space.
Our embedding layer has an input dimension of 273 (vocabulary size) and an output dimension of 64.

Following the embedding layer, the data are fed into the bidirectional Recurrent Neural Network (RNN) layer.
We used two RNN layers, each containing 64 units.
The effects of using Gated Recurrent Unit (GRU) or Long Short-Term Memory (LSTM)\cite{Hochreiter1997-vo} layers as the RNN layers were investigated (refer to Section~\ref{sec:GatedLayer}).
Using bi-RNN was motivated based on our previous work\cite{Ansari2023} in which LSTM helped improve the model's performance for predicting peptide properties using its sequences.
More details regarding RNN, LSTM, and GRU layers can be found in Ref. \citenum{Zhang2021-fa}.

The output from the bi-LSTM stack undergoes normalization via Layer Normalization\cite{Ba2016-dj}.
There is no agreement on why Layer Normalization improves the model's performance.\cite{Ioffe2015-gn, Awais2021-yt, Santurkar2018-nr, Xu2019-om}
The absence of a comprehensive theoretical understanding of normalization effects hinders the evolution of novel regularization schemes.\cite{Tian2022-wx}
Despite the limited understanding, Layer Normalization is employed due to its demonstrated effectiveness.\cite{Xu2019-om}

After normalization, data is processed through three dense layers containing 32, 16, and 1 units, respectively.
The 16-unit layer's output goes to two different 1-unit layers. 
One layer uses a linear function and the other uses a softplus function, producing $\hat{\mu}_m$ and $\hat{\sigma}_m$, respectively.

Negative log-likelihood loss $l$ was used to train the model.
It is defined as the probability of observing the label $y$ given the input $\vec{x}$:

\begin{equation}
    l(\vec{x},y) = \frac{log(\hat{\sigma}^2_m(\vec{x}))}{2} + \frac{\left(y-\hat{\mu}_m(\vec{x})\right)^2}{2\hat{\sigma}^2_m(\vec{x})}
    \label{eq:loss}
\end{equation}

During the training phase, dropout layers with 0.35 dropout rate were incorporated after the embedding and each dense layer to mitigate over-fitting.\cite{Gal2016-jc}
Models were trained using the Adam\cite{Kingma2014-hc} optimizer with a fixed learning rate of 0.0001 and default values for $\beta_1$ and $\beta_2$ (0.9 and 0.999, respectively).

Our model employs adversarial training, following the approach proposed by \citet{Lakshminarayanan2016-ck} to improve the robustness of our model predictions.
Because the input for our model is a discrete sequence, we generate adversarial examples by modifying the embedded representation of the input data.
Each iteration in the training phase consists of first computing the loss using Equation~\ref{eq:loss} and a second step with a new input $\vec{x}'$ to smooth the model's prediction:

\begin{equation}
    \vec{x}^\prime = \vec{x} + \epsilon \textrm{sign}(\nabla_x l(\vec{x},y))
\end{equation}
where $\epsilon$ is the strength of the adversarial perturbation.

Details of the model performance, limitations, training data, ethical considerations, and caveats are available as a model card\cite{Mitchell2019-nc} at \url{http://mol.dev/}.

\section{Results}\label{sec:Results}

In order to evaluate the performance of our model using deep ensembles, two baseline models were created: ($i$) an XGBoost Random Forest (RF) model using the 17 descriptors available on AqSolDB plus 1809 molecular descriptors calculated by PaDELPy, a python wrapper for the PaDEL-Descriptor\cite{Yap2011-ek} software, and ($ii$) a model with the same architecture used on our deep ensemble using RMSE as the loss function and no ensemble (referred to as DNN).
RFs are the state-of-the-art (SOTA) of solubility prediction. We used this baseline as a comparison to prove that our model is able to achieve SOTA performance using only molecular string representations.
In addition, we evaluate the effects of ($i$) the bi-RNN layer (either GRU or LSTM), ($ii$) using an augmented dataset to train, ($iii$) the adversarial training, and ($iv$) the ensemble size in the model's performance.
Table~\ref{tab:ResultsChallenge} shows the performance of each one of our trained models.

\begin{table}[ht]
\centering
\begin{tabular}{l|ccc|ccc|ccc}
        &\multicolumn{3}{|c}{Solubility Challenge 1} &   \multicolumn{3}{|c}{Solubility Challenge 2\_1} &   \multicolumn{3}{|c}{Solubility Challenge 2\_2}\\
Model   & RMSE & MAE  & r  & RMSE & MAE & r & RMSE & MAE  & r \\
\hline
RF
&1.121&0.914&0.547&\b{0.950}&\b{0.727}&\b{0.725}&\b{1.205}&\b{1.002}&\b{0.840}  \\
DNN
&1.540&1.214&0.433&1.315&1.035&0.651&1.879&1.381&0.736\\
DNN$_{Aug}$&
1.261&  1.007&  0.453& 1.371&  1.085&  0.453& 2.189&  1.710&  0.386\\
kde4$^{GRU}$&
1.610&  1.145& 0.462&  1.413&  1.114& 0.604& 1.488&   1.220&  0.704\\
kde4$^{LSTM}$&
1.554&  1.191& 0.507&  1.469&  1.188& 0.650& 1.523&   1.161&  0.706\\
kde4$^{GRU}$-NoAdv&
1.729&  1.348& 0.525&  1.483&  1.235& 0.622& 1.954&   1.599&  0.517\\
kde4$^{LSTM}$-NoAdv&
1.425&  1.114& 0.505&  1.258&  0.972&  0.610& 1.719&   1.439&  0.609\\
kde4$^{GRU}_{Aug}$&
1.329&  1.148& 0.426&  1.354&  1.157& 0.674& 1.626&   1.340&   0.623\\
kde4$^{LSTM}_{Aug}$&
1.273&  0.984&  0.473&  1.137&  0.932&  0.639&  1.511&  1.128&  0.717\\
kde8$^{LSTM}_{Aug}$&
1.247&  0.984& 0.542&  1.044&  0.846&  0.701&  1.418&  1.118&  0.729\\ 
kde10$^{LSTM}_{Aug}$-NoAdv&
1.689&  1.437& 0.471&  1.451&  1.238& 0.676& 1.599&   1.405&  0.699\\
kde10$^{LSTM}_{Aug}$&
\b{1.095}&  \b{0.843}&  \b{0.559}&  0.983&  0.793&  0.724&  1.263&  1.051&  0.792\\
\end{tabular}
\caption{Summary of the metrics for each trained model.
We used the Root Mean Squared Error(RMSE($\downarrow$)), Mean Absolute Error (MAE($\downarrow$)), and Pearson correlation coefficient (r($\uparrow$)) to evaluate our models.
The arrows indicate the direction of improvement.
Deep ensemble models are referred to as ``kde$N$'', where $N$ is the ensemble size.
Baseline models using random forest (RF) and the DNN model employed for deep ensemble (DNN) are also displayed.
DNN model was trained as described in Section \ref{sec:Methods}.
The models in which data augmentation was used were subscribed with the flag $Aug$.
A superscript indicates if the bidirectional layer implements a $GRU$ or a $LSTM$ layer.
In addition, models trained not using adversarial perturbation are flagged with ``-NoAdv''.
The columns show the results of each model evaluated on each solubility challenge dataset.
2\_1 represents the tight dataset (set-1), while 2\_2 represents the loose dataset (set-2) as described in the original paper (See Ref. \citenum{Llinas2019-eu}).
$r$ stands for the Pearson correlation coefficient.
The best-performing model in each dataset is displayed in bold.
}
\label{tab:ResultsChallenge}
\end{table}

\subsection{Gated layer}\label{sec:GatedLayer}

The most common RNN layers are the GRU and the LSTM.
GRU layers use two gates, reset and update, to control the cell's internal state.
On the other hand, LSTM layers use three gates: forget, input, and output, with the same objective.
Available studies compare GRU and LSTM performances in RNNs for different applications, for instance: forecasting\cite{Gao2020-kb}, cryptocurrency\cite{Kim2021-qj, Encean2022-br}, wind speed\cite{Kumar2021-se, Liu2021-jj}, condition of a paper press\cite{Mateus2021-nn}, motive classification in thematic apperception tests\cite{Gruber2020-nb} and music and raw speech\cite{Chung2014-vj}.
Nevertheless, it is not clear which of those layers would perform better at a given task.

We trained models with four elements in the deep ensemble using GRU or LSTM.
Metrics can be found in Table~\ref{tab:ResultsChallenge}; for an explanation of the naming syntax used in this work, refer to Table~\ref{tab:ResultsChallenge} caption.
Using LSTM resulted in a decrease in RMSE and MAE and an increase in the correlation coefficient, indicating better performance.
For Solubility Challenges 1, 2\_1, and 2\_2, the kde4$^{GRU}_{Aug}$ model yielded RMSE values of 1.329, 1.354, and 1.626, respectively, while the kde4$^{LSTM}_{Aug}$ model achieved 1.273, 1.137, and 1.511, respectively.
This trend was also observed for the models trained without data augmentation (See Table \ref{tab:ResultsChallenge}).
Considering that LSTM performs better regarding this model and data, we will consider only bi-LSTM layers for further discussion.
Those results are in accordance with our previous work\cite{Ansari2023} in which using LSTM helped improve the model's performance.

\subsection{Data augmentation}

Our model is not intrinsically invariant with respect to the SELFIES representation input.
For instance, both ``C(C(C1C(=C(C(=O)O1)O)O)O)O'' and ``O=C1OC(C(O)CO)C(O)=C1O'' are valid SMILES representations for the ascorbic acid (See Figure \ref{fig:model}) that will be encoded for different SELFIES tokens.
Hence, the model should learn to be invariant concerning changes in the string representation during training.
It can be achieved by augmenting the dataset with SMILES randomization and training the model using different representations with the same label.
Therefore, the model can learn relations in the chemical space instead of correlating the label with a specific representation.\cite{Arus-Pous2019-nl}
With this aim, we evaluated the effects of augmenting the dataset by generating new randomized SMILES representations for each sample.

Augmenting the dataset had a significant impact on the metrics.
It could be seen improvements of $\sim0.5$ in the RMSE when evaluating on challenge datasets 1 and 2\_1, and a gain of $\sim0.2$ on 2\_2 (See Table \ref{tab:ResultsChallenge}).
Concerning the first two datasets, augmenting data improved every model used in this study.
However, surprisingly, data augmentation led to a deprecation of the DNN model on the solubility challenge 2\_2 dataset.
This behavior 
was not further investigated.

\subsection{Adversarial training}

Using adversarial training improved performance in Lakshminarayanan \textit{et al.}\cite{Lakshminarayanan2016-ck} studies.
Hence, they suggested that it should be used in future applications of their deep learning algorithm.
Thus, we tested the effects of adversarial perturbation on training models with ensemble sizes of 4 and 10.

Comparing kde4$^{LSTM}$-NoAdv and kde4$^{LSTM}$, using adversarial training decreases model performance.
It can be seen in Table \ref{tab:ResultsChallenge} that using adversarial perturbation increased the RMSE from $1.425$ to $1.554$ and $1.258$ to $1.469$ in solubility challenges dataset 1 and 2\_1, respectively.
However, the RMSE decreased from $1.719$ to $1.523$ in dataset 2\_2.
Using adversarial perturbation affected our kde4$^{LSTM}$'s performance by a change in RMSE of $\pm0.2$.

The inconsistent performance improvement observed when using adversarial training was further investigated with models in which the dataset was augmented.
Due to the lack of multiple string representations in the training dataset, it is known that kde4$^{LSTM}$ may have generalization problems.
A generalization issue could direct the adversarial perturbation in a non-physical direction because the model does not have complete knowledge about the chemical representation space.
This hypothesis is reinforced when we compare kde10$^{LSTM}_{Aug}$-NoAdv and kde10$^{LSTM}_{Aug}$.
When using adversarial training on a model trained with an augmented dataset, the performance improvement is more evident ($\sim0.5$) and consistent for all the test datasets.

\subsection{Deep ensemble size}


To investigate the effects of increasing the ensemble size, we trained models with an ensemble of 4, 8, and 10 models.
Given the previous results, these models used LSTM as the bi-RNN layer and were trained on the augmented dataset.
Specifically for the solubility challenge 2\_2, the most complex set to predict, these models presented an RMSE of $1.511$, $1.418$, and $1.263$, respectively.
Therefore, increasing the ensemble size consistently improved performance.
We also observed this improvement on the other datasets (See Table \ref{tab:ResultsChallenge}).

\begin{figure}[h!]
    \centering
    kde4$^{LSTM}_{Aug}$\\
    \includegraphics[width=0.8\textwidth]{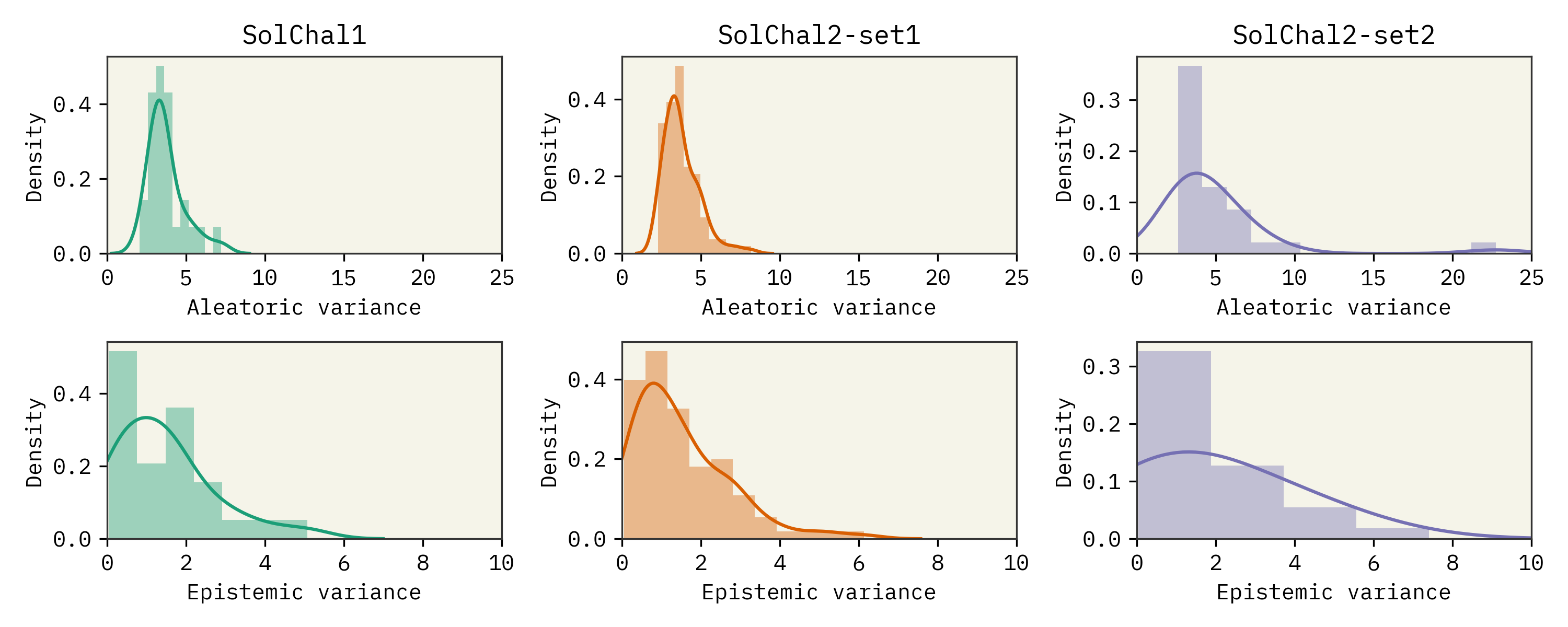}\\
    kde10$^{LSTM}_{Aug}$\\
    \includegraphics[width=0.8\textwidth]
    {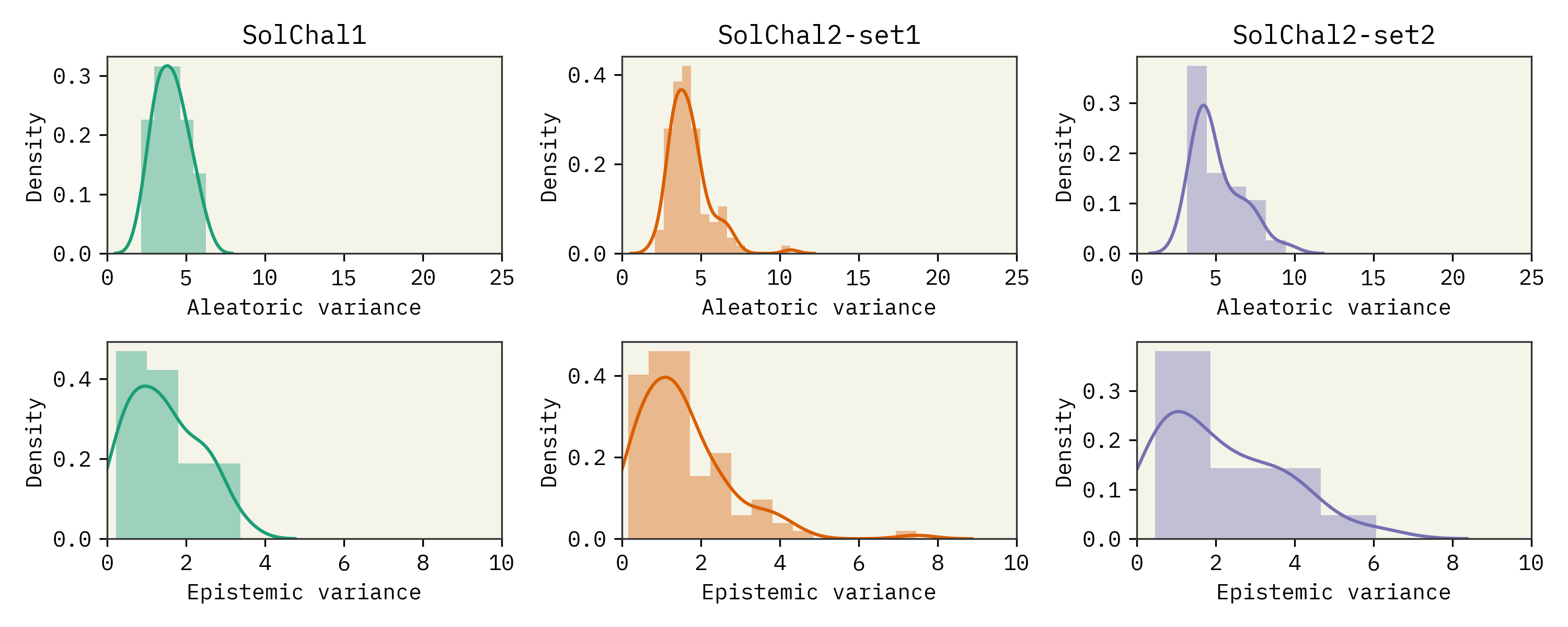}\\
    \caption{Density distribution of the aleatoric (AU) and epistemic variances (EU) for the: ($i$) kde4$^{LSTM}_{Aug}$ (top six panels) and ($ii$) kde10$^{LSTM}_{Aug}$ (bottom six panels).
    Increasing ensemble size reduces the extent of the distribution's tail, decreasing uncertainty about predictions.
    However, the ensemble size does not noticeably affect the distribution center.
    }
    \label{fig:uncertainty}
\end{figure}

Besides the immediate improvement in RMSE, increasing the ensemble size also improves the uncertainty of the model.
Figure \ref{fig:uncertainty} shows the density distribution of the aleatoric variance and the epistemic variance (respectively related to AU and EU) for kde4$^{LSTM}_{Aug}$ (top 6 panels) and kde10$^{LSTM}_{Aug}$ (bottom six panels).

The increase in ensemble size led to a decrease in both uncertainties.
AU distributions for the kde4$^{LSTM}_{Aug}$ are centered around 4 logS$^2$ , displaying a long tail that extends to values as high as 20 logS$^2$ in the worst case (solubility challenge 2\_2).
A similar trend is observed in EU distributions.
On the other hand, the kde10$^{LSTM}_{Aug}$ model results in narrower distributions.
The mean of these distributions remains relatively unchanged, but a noticeable reduction in the extent of their tails can be observed.
AU distribution ends in values around 10 logS$^2$.

\section{Discussion}
\label{sec:Discussions}

\begin{table}[ht]
\centering
\begin{tabular}{l|cc|cc|cc|cc}
& \multicolumn{2}{c|}{SolChal1} & \multicolumn{2}{c|}{SolChal2\_1} & \multicolumn{2}{c|}{SolChal2\_2} & \multicolumn{2}{c}{ESOL} \\
Model & RMSE & MAE & RMSE & MAE & RMSE & MAE & RMSE & MAE \\
\hline
RF                  & 1.121 & 0.914 & \b{0.950} & \b{0.727} & \b{1.205} & \b{1.002} & & \\
DNN                 & 1.540 & 1.214 & 1.315 & 1.035 & 1.879 & 1.381 & & \\
DNN$_{Aug}$         & 1.261 & 1.007 & 1.371 & 1.085 & 2.189 & 1.710 & & \\
kde4$^{LSTM}_{Aug}$ & 1.273 & 0.984 & 1.137 & 0.932 & 1.511 & 1.128 & 1.397 & 1.131 \\
kde8$^{LSTM}_{Aug}$ & 1.247 & 0.984 & 1.044 & 0.846 & 1.418 & 1.118 & 1.676 & 1.339 \\
kde10$^{LSTM}_{Aug}$& \b{1.095} & \b{0.843} & 0.983 & 0.793 & 1.263 & 1.051 & \b{1.316} & \b{1.089} \\
\hline
Linear regression\cite{Delaney2004-de} & & & & & & & 0.75 & \\
UG-RNN\cite{Lusci2013-rp}               & 0.90 & 0.74 & & & & & & \\
RF w/ CDF descriptors\cite{McDonagh2014-bj} & 0.93 & & & & & & & \\
RF w/ Morgan fingerprints\cite{Tayyebi2023-ad} & & \b{0.64} & & & & & & \\
Consensus\cite{Boobier2017-xk}          & \b{0.91} & & & & & & & \\
GNN\cite{Panapitiya2022-yo}             & $\sim1.10$ & & \b{0.91} & & \b{1.17} & & & \\
SolvBert\cite{Yu2022-gk}                & 0.925 & & & & & & & \\
SolTranNet$^a$\cite{Francoeur2021-zr}  & & & 1.004 & & 1.295 & & 2.99 & \\
SMILES-BERT$^{b}$\cite{Kim2021-yh}      & & & & & & & 0.47 & \\
MolBERT$^b$\cite{Fabian2020-vr}         & & & & & & & 0.531 & \\
RT$^{b}$\cite{Born2022-yt}            & & & & & & & 0.73 & \\
MolFormer$^b$\cite{Ross2022-na}         & & & & & & & \b{0.278} & \\
\end{tabular}
\caption{Metrics for the best models found in the current study (upper section) and for other state-of-art models available in the literature (lower section).
Values were taken from the cited references.
Missing values stand for entries that the cited authors did not study.
SolChal columns stand for the Solubility Challenges. 
2\_1 represents the tight dataset (set-1), while 2\_2 represents the loose dataset (set-2) as described in the original paper (See Ref. \citenum{Llinas2019-eu}). The best-performing model in each dataset has its RMSE value in bold.\\
$^a$ Has overlap between training and test sets.\\
$^b$ Pre-trained model was fine-tuned on ESOL.\\
}
\label{tab:results}
\end{table}

\begin{figure}[ht]
    \centering
    kde4$^{LSTM}_{Aug}$\\
    \includegraphics[width=0.9\textwidth]{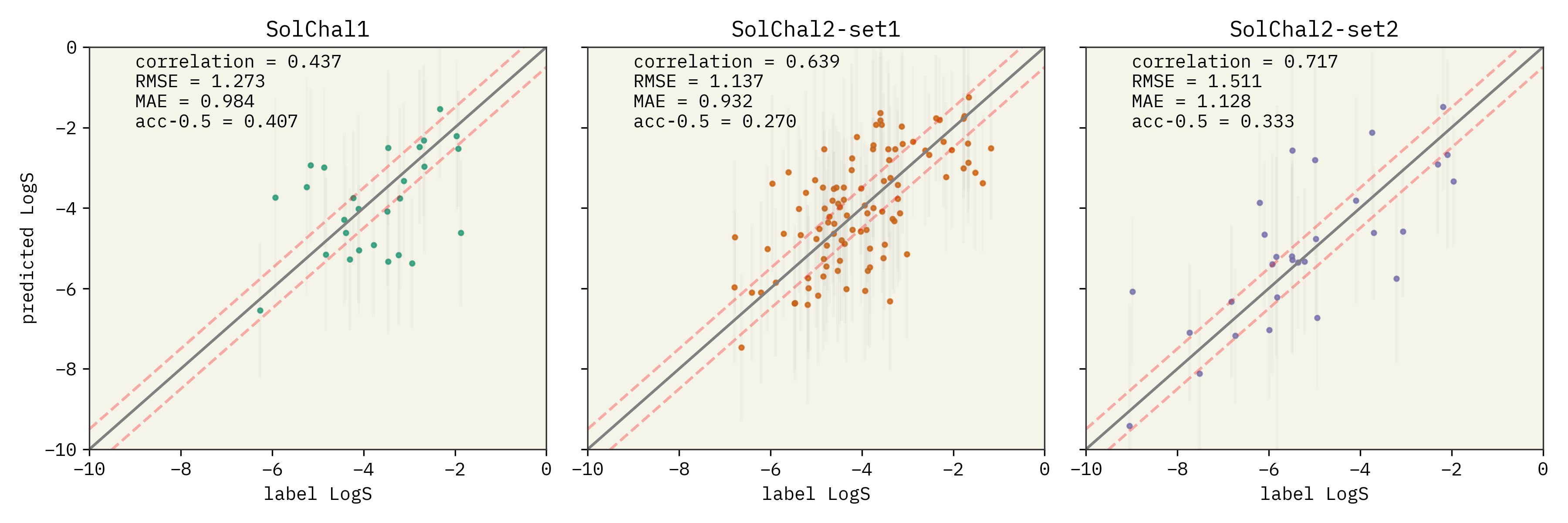}\\
    kde10$^{LSTM}_{Aug}$\\
    \includegraphics[width=0.9\textwidth]{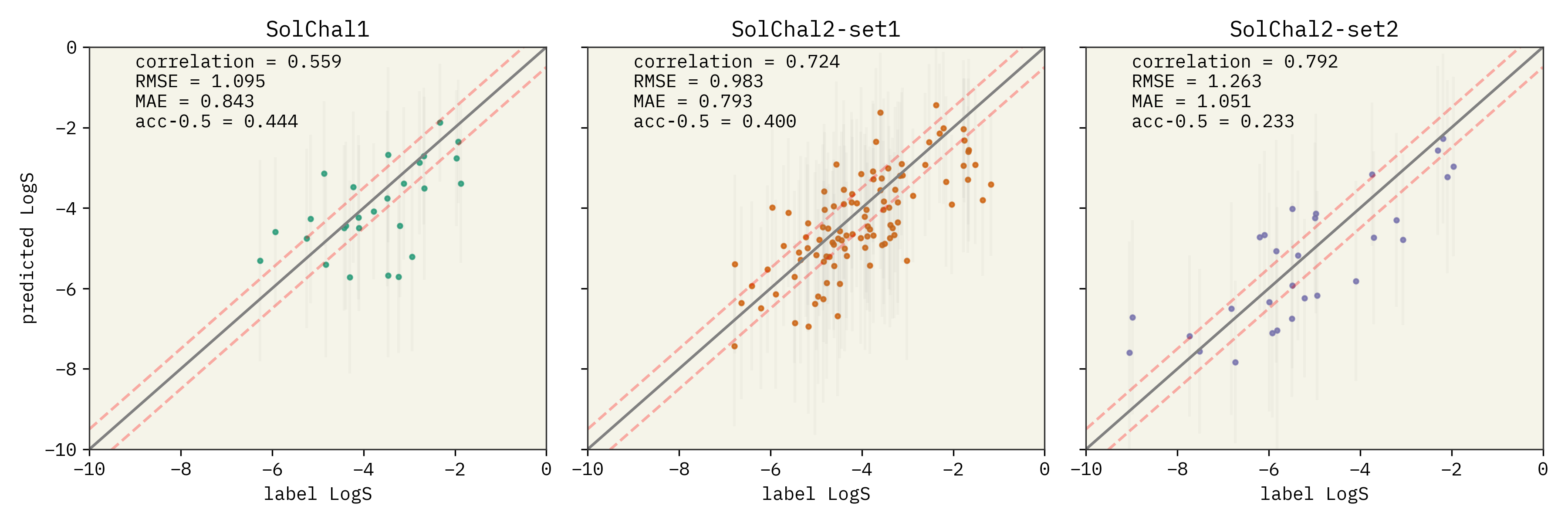}
    \caption{Parity plots for two selected models being evaluated on the solubility challenge datasets: 
    ($i$) kde4$^{LSTM}_{Aug}$ (top row), 
    and ($ii$) kde10$^{LSTM}_{Aug}$ (botom row).
    The left, middle, and right columns show the parity plots for solubility challenge 1\cite{Llinas2008-rc}, 2-set1, and 2-set2\cite{Llinas2019-eu}, respectively.
    Pearson correlation coefficient is displayed together with RMSE and MAE.
    ``acc-0.5'' stands for the $\pm0.5\textrm{log}\%$ metric.
    Red dashed lines show the limits for molecules considered a correct prediction when computing the $\pm0.5\textrm{log}\%$.
    The correlation between predicted values and labels increases when more models are added to the ensemble.
    RMSE and MAE also follow this pattern.
    However, the $\pm0.5\textrm{log}\%$ decreases in set-2 of the second solubility challenge dataset (SolChal2-set2).
    While kde10$^{LSTM}_{Aug}$ improved the prediction of molecules that were being poorly predicted by kde4$^{LSTM}_{Aug}$, the prediction of molecules with smaller errors was not greatly improved.}
    \label{fig:models}
\end{figure}

After extensively investigating the hyperparameter selection, we compared our model with available state-of-art models from the literature.
Performance metrics on the solubility challenge datasets can be found in Table \ref{tab:results}. 
Parity plots for our chosen models are presented in Figure \ref{fig:models}.

Comparing the performance of different models is a complex task, as performance metrics cannot be directly compared across models evaluated on distinct datasets. 
To address this issue, \citet{Panapitiya2022-yo} curated a large and diverse dataset to train models with various architectures and molecular representations. 
They also compared the performance of these models on datasets from the literature\cite{Yalkowsky1992-zc, Huuskonen2000-pb, Klopman2001-hk, Hou2004-bw, Delaney2004-de, Wang2007-ub, Llinas2008-rc, Boobier2017-xk, Llinas2019-eu, Boobier2020-mc, Tang2020-mz, Cui2020-zh}.
Although their models achieved an RMSE of $\sim1.1$ on their test set, using descriptors as molecular representations resulted in RMSE values ranging from $0.55$ to $\sim1.35$ when applied to other datasets from the literature.
According to their study, the Solubility Challenge datasets by Llinàs et al.\cite{Llinas2008-rc, Llinas2019-eu} were found to be particularly challenging due to their more significant reproducibility error.
Therefore, we focused on the Llinàs datasets to compare our performance with the literature.

Focusing on the solubility challenge 1 dataset\cite{Llinas2008-rc},
kde10$^{LSTM}_{Aug}$ is only $\sim0.2$ RMSE units worse than the best model available in the literature\cite{Lusci2013-rp}.
The RMSE of the participants of the challenge was not reported.\cite{Hopfinger2009-rt}
The primary metric used to evaluate models was the percentage of predictions within an error of 0.5 LogS units (called $\pm0.5\textrm{log}\%$).
Computing the same metric, kde10$^{LSTM}_{Aug}$ has a percentage of correct prediction of 44.4\%.
This result would place our model among the 35\% best participants.
The participant with the best performance presented a $\pm0.5\textrm{log}\%$ of 60.7\%.

The architecture of the models was not published in the findings of the first challenge.\cite{Hopfinger2009-rt}
Nevertheless, the findings for the second challenge\cite{Llinas2020-ea} investigated the participants more thoroughly.
Participants were asked to identify their models' architecture and descriptors used.
The challenge is divided into two datasets.
Set-1 contains LogS values with an average interlaboratory reproducibility of 0.17 LogS.
Our kde10$^{LSTM}_{Aug}$ achieve an RMSE of 0.983 and a $\pm0.5\textrm{log}\%$ of 40.0\% in this dataset.
Therefore, our model performs better than 62\% of the published RMSE values and 50\% of the $\pm0.5\textrm{log}\%$.
In addition, the model with the best performance is an artificial neural network (ANN) that correctly predicted 61\% ($\pm0.5\textrm{log}\%$) of the molecule's LogS using a combination of molecule descriptors and fingerprints.
The second dataset (set-2) contains molecules whose solubility measurements are more challenging, reporting an average error in reproducibility of 0.62 LogS.
The kde10$^{LSTM}_{Aug}$ achieves an RMSE of 1.263 and a $\pm0.5\textrm{log}\%$ of 23.3\%.
It performs better than 82\% of the candidates when considering the RMSE.
Surprisingly, $\pm0.5\textrm{log}\%$ does not follow this outstanding performance, which is more significant than only 32\% 
Regarding the literature, kde10$^{LSTM}_{Aug}$ has an RMSE only $\sim0.1$ higher than a GNN that used an extensive set of numeric and one-hot descriptors in their feature vector.\cite{Panapitiya2022-yo}
Our model performs better than a transformer model that uses SMILES-string and an adjacency matrix and inputs.\cite{Francoeur2021-zr}
The performance of those models is available in Table \ref{tab:results}.

Notably, all participants in the solubility challenge 2 submitted a kind of QSPR or descriptor-based ML model.
Using descriptors provides an easy way to ensure model invariance concerning molecule representation and is more informative since they can be physical quantities.
However, selecting appropriate descriptors is crucial for developing descriptor-based ML models.
It often requires specialists with a strong intuition about the relevant physical and chemical properties for predicting the target quantity.
Feature-based models are still being considered to be the SOTA of solubility prediction. Recently, studies investigating different descriptors and fingerprints were performed.\cite{Tayyebi2023-ad, Zagidullin2021-gw} These studies showed that similarly to the impacts of data quality\cite{Sorkun2021-kc}, molecular representation also has a great impact on models' performance. Despite \citet{Tayyebi2023-ad} being able to achieve an MAE of 0.64 on solubility challenge 1 when using Morgan fingerprints (MF), \citet{Zagidullin2021-gw} reported poor performance when using MF.
Our approach, on the other hand, is based on extracting information from simple string representations, a more straightforward raw data.
Furthermore, we could achieve SOTA performance while balancing the model size and complexity and using a raw input (a simple string).
This simplified usage enables running the model on devices with limited computing power.

Lastly, transformer models have been used to address the issue of accurately predicting the solubility of small compounds.
The typical workflow for transformers involves pre-training the model using a large dataset and subsequently fine-tuning it for a specific downstream task using a smaller dataset.
Most existing models were either pre-trained on the ESOL\cite{Delaney2004-de} dataset or pre-trained on a larger dataset and fine-tuned using ESOL.
Hence, the generalizability of those models cannot be verified.
In a study by \citet{Francoeur2021-zr}, they considered two versions of their model, SolTranNet.
The first version of SolTranNet was trained with the ESOL dataset using random splits.
This approach achieved an RMSE of 0.278.
Subsequently, the deployed version of SolTranNet was trained with the AqSolDB\cite{Sorkun2019-jv}.
When ESOL was used to evaluate their deployed version, the model presented an RMSE of 2.99.
While our model achieved an RMSE of 1.316 on ESOL, outperforming the SolTranNet deployed version, it cannot be compared with other models trained on ESOL.

\section{Conclusions}

We used TensorFlow.js, the JavaScript library of TensorFlow, to develop a deep ensemble recurrent neural network (RNN) capable of accurately predicting LogS values from SMILES or SELFIES string representations. This model is accessible via a static website at \url{https://mol.dev/}. The significant contributions of this research include: (1) Demonstrating the feasibility of using string representations for solubility prediction; (2) establishing that using string representations does not significantly compromise performance, with our models achieving results comparable to state-of-the-art (SOTA) models on the datasets by Llinas et al.; (3) enhancing prediction reliability and practical applicability by incorporating uncertainty estimates; and (4) significantly facilitating model usability through the deployment on a static website, eliminating the need for domain-specific data or expertise.

Our based on a deep ensemble of recurrent neural networks (RNNs) model was trained using SMILES randomization for data augmentation on the AqSolDB dataset and validated using the solubility challenges by Llínas et al. It directly processes molecular string representations, such as SMILES or SELFIES, to predict solubility without relying on pre-selected descriptors.
This approach not only simplifies the prediction process but also enhances its applicability across a broader chemical space.
In addition, we show that this deep ensemble RNN model could achieve similar performance compared to a random forest (RF) using PaDEL descriptors. RFs with descriptors were shown to perform relatively well in other datasets.

By carefully compromising between performance and complexity, we developed a model with acceptable performance and that is not computationally intensive. It enables us to host the model on a static website using TensorFlow JS.
Our model was designed to operate on devices with limited computational resources, aiming to broaden the accessibility of advanced solubility prediction tools. 
This application can satisfactorily run on any device with limited computational resources, such as laptops and smartphones.
This approach ensures wider applicability, catering to the needs of users without access to high-performance computing facilities, improving usability and flexibility, and decreasing implementation costs.
We believe this is a considerable step in improving the usability of deep learning models and promoting such models to a broader scientific community.

\section*{Data and code availability}
All code needed to reproduce those results is publicly available on the following GitHub repository: \url{https://github.com/ur-whitelab/mol.dev}.
The model is also publicly accessible at the following address: \url{https://mol.dev/}.

\section*{Authors contribution}
M.C.R. implemented the deep learning model, performed the training and hyperparameters optimization, tested the model's performance, analyzed the results, and wrote this manuscript.
A.D.W. idealized the project, proposed the model to be used, implemented the deep learning approach, and developed \url{http://mol.dev/}.

\subsection*{Conflicts of interest}
The authors have no conflicts to declare.

\subsection*{Acknowledgments}
The authors acknowledge the National Institute of General Medical Sciences of the National Institutes of Health (NIH) under award number R35GM137966.
This research used the computational resources and structure provided by the Center for Integrated Research Computing (CIRC) at the University of Rochester.

\bibliographystyle{unsrtnat}
\bibliography{refs}

\end{document}